\documentstyle[epsfig,12pt]{article}
\newcommand{\titul}[1] {\begin{center}{\Large {\bf #1 } } \end{center}
\vskip 0.8cm}

\newcommand{\autor}[1] {\begin{center}  {\bf \lineskip .3cm #1  }
                        \end{center} }

\newcommand{\lugar}[1] {\begin{center}  {\normalsize \bf \it #1   }
\end{center}}
\begin{document}
\hbadness=10000
\pagenumbering{arabic}
\begin{titlepage}
%


%
\titul{\bf Resolution to the $B\to\phi K^*$ polarization puzzle}
\autor{Hsiang-nan Li\footnote{E-mail: hnli@phys.sinica.edu.tw} }

\lugar{Institute of Physics, Academia Sinica, Taipei, Taiwan 115,
Republic of China}
\lugar{Department of Physics, National Cheng-Kung University,\\
Tainan, Taiwan 701, Republic of China}

\vskip 2.0cm

\thispagestyle{empty}
\vspace{10mm}
\begin{abstract}

We resolve the $B\to\phi K^{*}$ polarization puzzle by postulating
a smaller $B\to K^*$ form factor $A_0\approx 0.3$ and by adding
penguin annihilation and nonfactorizable contributions from the
perturbative QCD approach. If this explanation is valid, the
penguin-dominated modes governed by the $B\to K^*$ form factors,
such as $B^+\to K^{*+} K^{*0}$ and $B^0\to K^{*0} {\bar K}^{*0}$,
should exhibit similar polarization fractions. Our resolution is
compared with others in the literature, and experimental
discrimination is proposed.

\end{abstract}
\thispagestyle{empty}
\end{titlepage}


To understand the polarization fractions of the $B\to\phi K^*$
decays has been a challenge. Motivated by this subject, we have
investigated most of the $B\to VV$ modes, and observed that they
are classified into four categories \cite{LM04}. First, the
$B^0\to (D_s^{*+}, D^{*+}, \rho^+)D^{*-}$ modes can be understood
by kinematics in the heavy-quark limit, whose longitudinal
polarization fractions $R_L\sim 0.5$, $0.5$, and $0.9$
\cite{PDG,Belich}, respectively, follow the mass hierarchy among
the $D_s^{*}$, $D^{*}$ and $\rho$ mesons emitted from the weak
vertex. Second, the $B\to (\rho, \omega)\rho$ modes are understood
by kinematics in the large-energy limit, leading to $R_L\sim 1$
\cite{BelleRhopRho0,gritsan,Bar0411}. Applying the same
estimation, we have predicted $R_L\sim 0.7$ for the $B^+\to
(D_s^{*+}, D^{*+})\rho^0$ decays, which can be compared with
future data. For penguin-dominated modes, such as those listed in
Table~\ref{tab:tab1}, the polarization fractions deviate from the
naive counting rules based on kinematics \cite{CKL2}: the
annihilation contribution from the $(S-P)(S+P)$ operators and the
nonfactorizable contribution decrease $R_L$ to about 0.75 for the
pure-penguin $B^+\to\rho^+ K^{*0}$ decay. Adding a tree
contribution, $R_L$ of $B^+\to\rho^0 K^{*+}$ can go up to about
0.9 \cite{LM04}. The fourth category, consisting of the puzzling
$B\to\phi K^*$ decays, is also pure-penguin, but its $R_L\sim 0.5$
shown in Table~\ref{tab:tab1} is much lower than 0.75.

\begin{table}[ht]
\begin{center}
\begin{tabular}{c c c c}\hline \hline
Mode&Pol. Fraction&Belle&Babar \\
\hline
$B^+\to\phi K^{*+}$&$R_L$&$0.49\pm 0.13\pm 0.05$ \cite{Zhang04}&$0.46\pm0.12\pm0.03$ \cite{BaBarRhopRho0}\\
&$R_\perp$&$0.12^{+0.11}_{-0.08}\pm 0.03$ \cite{Zhang04}&\\
$B^0\to\phi K^{*0}$&$R_L$&$0.52\pm 0.07\pm 0.05$ \cite{Zhang04} 
&$0.52\pm0.05\pm0.02$ \cite{Bar017}\\
&$R_\perp$&$0.30\pm 0.07\pm 0.03$ \cite{Zhang04}
&$0.22\pm0.05\pm0.02$ \cite{Bar017}\\
\hline \hline
Mode&Pol. Fraction&Belle&Babar\\
\hline
$B^+\to\rho^0 K^{*+}$&$R_L$&&$0.96^{+0.04}_{-0.15}\pm0.04$ \cite{BaBarRhopRho0}\\
$B^+\to\rho^+ K^{*0}$&$R_L$&$0.50\pm 0.19^{+0.05}_{-0.07}$ \cite{Bel102}&$0.79\pm 0.08\pm0.04\pm 0.02$ \cite{Bar093}\\
\hline \hline
\end{tabular}
\end{center}
\caption{Polarization fractions in the penguin-dominated $B\to VV$
decays. }\label{tab:tab1}
\end{table}

It seems that the $B\to \phi K^*$ polarizations are the only
anomaly so far, and many attempts to resolve it have been
proposed, which include new physics \cite{G03,YWL}, the
annihilation contribution \cite{AK} in the framework of
QCD-improved factorization (QCDF) \cite{BBNS}, the charming
penguin in soft-collinear effective theory (SCET) \cite{BPRS}, the
rescattering effect \cite{CDP,LLNS,CCS}, and the $b\to sg$
transition (the magnetic penguin) \cite{HN}. We have carefully
analyzed these proposals \cite{LM04}: the annihilation amplitude
has to be parameterized in QCDF, and varying free parameters to
fit the data can not be conclusive \cite{Li04}. The charming
penguin strategy, demanding many free parameters, does not help
understand dynamics. Moreover, it has been argued that the
charming penguin, without infrared divergences from diagrammatic
calculations, should be factorizable in the current leading-power
SCET formalism \cite{LM04}. A similar criticism has been raised
recently in \cite{BBNS04}. The rescattering effect is based on a
model-dependent analysis \cite{W,Ligeti04}, and constrained by the
$B\to\rho K^{*}$ data. The prediction $R_\parallel\gg R_\perp$ for
$B\to\phi K^*$ \cite{CCS}, $R_\parallel$ and $R_\perp$ being the
parallel and perpendicular polarization fractions, respectively,
also contradicts the observed relation $R_\parallel\approx
R_\perp$ in Table~\ref{tab:tab1}. The exotic magnetic penguin is
suppressed by the $G$-parity, and not sufficient to reduce $R_L$
down to 0.5 \cite{LM04}. However, we are not claiming a signal of
new physics, since the complicated QCD dynamics in $B\to VV$
decays has not yet been fully explored.

In this letter we shall investigate whether QCD effects can
resolve the $B\to\phi K^*$ polarization puzzle without resorting
to exotic mechanism or new physics. These decays have been studied
in the perturbative QCD (PQCD) approach \cite{LY1,KLS,LUY}, and
the results of the branching ratios, the magnitudes of the
helicity amplitudes $A_{L}$, $A_{\parallel}$, and $A_{\perp}$, and
their relative strong phases $\phi_\parallel$ and $\phi_\perp$ are
summarized in Table~\ref{tab5} \cite{CKL2}. The normalization of
these amplitudes have been chosen, such that they satisfy
\begin{eqnarray}
|A_{L}|^2+|A_{\parallel}|^2+|A_{\perp}|^2=1\;,
\end{eqnarray}
with $|A_{L}|^2=R_L$, $|A_{\parallel}|^2=R_\parallel$, and
$|A_{\perp}|^2=R_\perp$. The first rows (I), coming only from the
factorizable emission topology, correspond to the results under
the factorization assumption (FA) \cite{BSW}. It is obvious that
the polarization fractions $R_L\approx 0.92$ and
$R_\parallel\approx R_\perp\approx 0.04$ follow the naive counting
rules,
\begin{eqnarray}
R_L\sim 1-O(m_\phi^2/m_B^2)\;,\;\;\;\;R_\parallel \sim R_\perp\sim
O(m_\phi^2/m_B^2)\;, \label{nai}
\end{eqnarray}
$m_B$ ($m_\phi$) being the $B$ ($\phi$) meson mass.

\begin{table}[htbp]
\begin{center}
\begin{tabular}{ccccccc}
\hline\hline
Mode & Br $(10^{-6})$ & $ |A_{L}|^{2}$ & $ |A_{\parallel}|^{2}$ &
$ |A_{\perp}|^{2}$ & $\phi_{\parallel}(rad.)$ &
$\phi_{\perp}(rad.)$
\\ \hline
$\phi K^{*0}$(I) &$14.48$ &$0.923$  & $0.040$ & $0.035$ &
$\pi$ & $\pi$\\
\hspace{0.7cm}(II)& $13.25$ &$0.860$ & $0.072$ & $0.063$ & $3.30$
& $3.33$ \\
\hspace{0.7cm}(III) & $16.80$ &$0.833$ & $0.089$ & $0.078$ &
$2.37$ & $2.34$ \\
\hspace{0.7cm}(IV) & $14.86$& {$0.750$} & $0.135$ & $0.115$ & $2.55$ & $2.54$ \\
\hline $\phi K^{*+}$(I)  & $15.45$ &$0.923$ & $0.040$ & $0.035$ &
$\pi$  & $\pi$  \\
\hspace{0.7cm}(II) &$14.17$ &$0.860$ & $0.072$ & $0.063$ & $3.30$
& $3.33$
\\
\hspace{0.7cm}(III) & $17.98$&$0.830$ & $0.094$ & $0.075$ &
$2.37$ & $2.34$ \\
\hspace{0.7cm}(IV)  & $15.96$&{$0.748$} & $0.133$ & $0.111$ & $2.55$ & $2.54$ \\
\hline
$\phi K^{*0}$&   $10.2^{+2.5}_{-2.1}$   & $0.59^{+0.02}_{-0.02}$
& $0.22^{-0.01}_{+0.01}$ & $0.19^{-0.01}_{+0.01}$ & $2.32^{+0.11}_{-0.13}$& $2.31^{+0.12}_{-0.13}$ \\
\hline\hline
\end{tabular}
\end{center}
\caption{(I) Without the nonfactorizable and annihilation
contributions, (II) add  only the nonfactorizable contribution,
(III) add only the annihilation contribution, and (IV) add both
the nonfactorizable and annihilation contributions. The last row
is for $A_0=0.28$.}\label{tab5}
\end{table}

The next-to-leading-power annihilation amplitudes, mainly from the
$(S-P)(S+P)$ operators, and the nonfactorizable amplitudes bring
the first rows into the fourth ones (IV) with the fractions
$R_L\approx 0.75$. We observe from the second and third rows, (II)
and (III), that these subleading corrections work toward the
direction indicated by the data. It is easy to understand the
sizable deviation from Eq.~(\ref{nai}) caused by these subleading
corrections, which are of $O(m_\phi/m_B)$ for all the three final
helicity states \cite{CKL2}. If they are of the same order of
magnitude as and constructive to the transverse polarization
amplitudes, an enhancing factor will be gained, which may be large
enough to modify the counting rules numerically (note that
$m_\phi/m_B$ is only about $1/5$). However, the total effect, as
shown in Table~\ref{tab5}, is not sufficient to lower $R_L$ of the
$B\to\phi K^*$ decays down to around 0.5. The branching ratios in
(I) and in (IV) are roughly equal, indicating that the subleading
corrections decrease the longitudinal components and increase the
transverse ones by roughly equal amount.

Two nice features exhibited in Table~\ref{tab5} are that PQCD has
predicted $R_\parallel \approx R_\perp$, contrary to those from
the rescattering effect \cite{CCS}, and that the relative strong
phases among the helicity amplitudes are consistent with the
$B^0\to\phi K^{*0}$ data:
\begin{eqnarray}
& &\phi_{\parallel}=2.21\pm 0.22\pm 0.05\,
(rad.)\;,\;\;\;\;\phi_{\perp}=2.42\pm 0.21 \pm
0.06\,(rad)\,\cite{Zhang04} \;,\nonumber\\
& &\phi_{\parallel}=2.34^{+0.23}_{-0.20}\pm 0.05\,
(rad.)\;,\;\;\;\;\phi_{\perp}=2.47\pm 0.25\pm
0.05\,(rad)\,\cite{Bar017}\;.\label{phase}
\end{eqnarray}
The former implies that the rescattering effect may not be
essential in $B$ meson decays into two light mesons \cite{CL00}.
The consistency of the predicted $\phi_\parallel$ and $\phi_\perp$
with the data, once again, supports that the evaluation of strong
phases in PQCD is reliable. Other examples include the predictions
for the direct CP asymmetries in the $B\to K^+\pi^-$, $\pi^+\pi^-$
modes \cite{KLS,LUY}, and the results of the $B\to
D^{(*)}\pi(\rho)$ branching ratios, which crucially depend on the
strong phases of the color-suppressed amplitudes.

As emphasized above, the $B\to \phi K^*$ polarizations are very
unique, and it is difficult to find new mechanism, which affects
only these modes but not others. Hence, we do not intend to
propose any new mechanism or new physics to resolve the puzzle. To
explain our idea, we quote the explicit expressions of the three
helicity amplitudes in terms of the $B\to K^*$ transition form
factors in FA \cite{LM04},
\begin{eqnarray}
A_{L}&\propto&2r_2\epsilon_2^*(L)\cdot \epsilon_3^*(L)A_0
\;, \label{al}\\
A_{\parallel}&\propto&-\sqrt{2}(1+r_2)A_1\;, \label{ap} \\
A_{\perp}&\propto&- \frac{2r_2 r_3}{1+r_2}\sqrt{2[(v_2\cdot
v_3)^{2}-1]} V\;, \label{ase1r}
\end{eqnarray}
with the $K^*$ ($\phi$) meson velocity $v_2$ ($v_3$) and
polarization vector $\epsilon_2$ ($\epsilon_3$), $r_2=m_{K^*}/m_B$
and $r_3=m_{\phi}/m_B$. The form factors $A_0$, $A_1$, and $V$ in
the standard definitions obey the symmetry relations in the
large-energy limit \cite{BF,Ch},
\begin{eqnarray}
&&\frac{m_B}{m_B+m_{K^*}} V = \frac{m_B+m_{K^*}}{2 E} A_1 = T_1
=\frac{m_B}{2E}T_2\;,
\label{rho1}\\
& &\frac{m_{K^*}}{E}A_0= \frac{m_B+m_{K^*}}{2 E}A_1 -
\frac{m_B-m_{K^*}}{m_B}A_2\;,
\label{rho2}
\end{eqnarray}
where $T_1$ and $T_2$ are the form factors involved in the $B\to
K^*\gamma$ decays, and $E$ is the $K^*$ meson energy.

The results in Table~\ref{tab5} correspond to the form factors
$A_0=0.40$, $A_1=0.26$ and $V=0.35$. First, the $B\to K^*\gamma$
branching ratios have constrained the form factors $T_1\approx
T_2\approx 0.3$ \cite{A02,BFS}, which are also in agreement with
the lattice result \cite{B02}. Compared to the symmetry relation
in Eq.~(\ref{rho1}), it is obvious that PQCD has given reasonable
values of $A_1$ and $V$. Second, there has not yet been any
measurement, except $B\to \phi K^*$, which constrains $A_0$. The
other penguin-dominated $B\to\rho(\omega)K^*$ decays are mainly
governed by the $B\to\rho(\omega)$ form factors. Third, the PQCD
predictions for the $B\to\phi K^*$ branching ratios in
Table~\ref{tab5} are larger than the data \cite{HFAG},
\begin{eqnarray}
B(B^0\to \phi K^{*0})=\left(9.5\pm 0.9\right)\times
10^{-6}\;,\;\;\;\;B(B^+\to \phi K^{*+})=\left(9.7\pm
1.5\right)\times 10^{-6}\;.\label{br}
\end{eqnarray}
Note that the same value of $A_0\approx 0.40$ leads to the
branching ratio about $10\times 10^{-6}$ for the longitudinal
component in PQCD, but about $5\times 10^{-6}$ in QCDF
\cite{AK,CY}, because of the dynamical penguin enhancement in the
former \cite{KLS}. The above three observations hint that the PQCD
results for the transverse components of the $B\to\phi K^*$ decays
should have been reasonable, and that the longitudinal components
may have been overestimated. We are then led to conjecture that a
smaller $A_0$ will resolve the puzzle, giving both lower $R_L$ and
lower branching ratios.

In PQCD, a $B\to K^*$ form factor is written as the convolution of
a hard kernel with the $B$ meson wave function and with a set of
$K^*$ meson distribution amplitudes. Note that the form factors
$A_0$, $A_1$ and $V$ involve different sets of $K^*$ meson
distribution amplitudes: the twist-2 $\phi_{K^*}$, and the
two-parton twist-3 $\phi_{K^*}^t$ and $\phi_{K^*}^s$ for $A_0$,
and the twist-2 $\phi_{K^*}^T$, and the two-parton twist-3
$\phi_{K^*}^v$ and $\phi_{K^*}^a$ for $A_1$ and $V$ (the notations
are referred to \cite{CKL2}). Our investigation indicates that the
latter set of model distribution amplitudes derived from QCD sum
rules \cite{BBKT} has been acceptable, but the former set has not.
Recently, the reanalysis of $\phi_{K^*}$, parameterized as
\begin{eqnarray}
\phi_{K^*}(x) &=&\frac{3f_{K^*}}{\sqrt{2N_{c}}}x(1-x)
\Big[1+3a_1^{K^*}(1-2x)\Big]\;,
\label{pk1}
\end{eqnarray}
showed that the Gegenbauer coefficient has a revised value
$a_1^{K^*}=0.10\pm 0.07$ \cite{BL04}, different from
$a_1^{K^*}=0.19\pm 0.05$ in \cite{BBKT}. That is, considering the
theoretical uncertainty, $\phi_{K^*}(x)$ could be quite close to
the asymptotic model corresponding to $a_1^{K^*}=0$.

To test our idea, we choose the asymptotic models for the $K^*$
meson distribution amplitudes relevant to the evaluation of $A_0$:
\begin{eqnarray}
\phi_{K^* }( x) &=&\frac{3f_{K^*}}{\sqrt{2N_{c}}}x(1-x)\;,
\label{pk2}\\
\phi_{K^*}^{t}( x) &=&\frac{f_{K^*}^T}{2\sqrt{2N_{c}}} 3(1-2x)^2
\;,
\label{pk3t}\\
\phi _{K^*}^s( x)  &=&\frac{f_{K^*}^T}{2\sqrt{2N_{c}}} 3(1-2x)\;,
\label{pk3s}
\end{eqnarray}
which lead to $A_0=0.28$, about 70\% of the original value. The
main reduction is caused by the change of $\phi _{K^*}^s$ in
Eq.~(\ref{pk3s}). The model-dependent evaluations of $A_0$ vary in
a wide range from 0.31 to 0.47, and $A_0\approx 0.3$ has been
supported by the recent covariant light-front QCD (LFQCD)
calculation \cite{CCH}. This smaller value does not contradict to
any existing data as emphasized above. We suggest to also
reanalyze $\phi_{K^*}^t$ and $\phi_{K^*}^s$ in QCD sum rules, so
that it is possible to examine whether a consistency between the
PQCD and LFQCD results of $A_0$ can be achieved.

The models for the distribution amplitudes $\phi_{K^*}^{T}$,
$\phi_{K^*}^{v}$ and $\phi_{K^*}^{a}$, relevant to the evaluation
of the form factors $A_1$ and $V$, and those for the $\phi$ meson
distribution amplitudes and for the $B$ meson wave function,
remain the same as in \cite{CKL2}. We then compute all amplitudes,
including the penguin emission, penguin annihilation and
nonfactorizable ones, for the longitudinal and transverse
polarizations using the $k_T$ factorization formulas in
\cite{CKL2}. The numerical outcomes are listed as the last row in
Table~\ref{tab5}. Simply adopting the asymptotic models in
Eqs.~(\ref{pk2})-(\ref{pk3s}), the modified branching ratio
$10.2\times 10^{-6}$, the polarization fractions $R_L=0.59$ and
$R_\parallel\approx R_\perp$, and the relative strong phases
$\phi_\parallel\approx \phi_\perp\approx 2.3$, are all consistent
with the $B^0\to\phi K^{*0}$ data in Table~\ref{tab:tab1}, and in
Eqs.~(\ref{phase}) and (\ref{br}). Therefore, we claim that the
measured $B\to\phi K^*$ polarizations might imply nothing but a
smaller form factor $A_0$, and that their explanation does not
require any exotic mechanism or new physics. The penguin
annihilation and nonfactorizable contributions play an important
role here. In FA without these contributions, $A_0$ has to be as
small as 0.15 in order to reach $R_L\sim 0.6$, for which the
$B\to\phi K^*$ branching ratios will fall far below the data.

The central values in the last row of Table~\ref{tab5} correspond
to the shape parameter $\omega_B=0.40$ GeV, appearing in the $B$
meson wave function \cite{KLS},
\begin{eqnarray}
\phi_B(x,b)=N_Bx^2(1-x)^2
\exp\left[-\frac{1}{2}\left(\frac{xm_B}{\omega_B}\right)^2
-\frac{\omega_B^2 b^2}{2}\right]\;,
\end{eqnarray}
where the normalization constant $N_B$ is related to the decay
constant $f_B$ through
\begin{eqnarray}
\int dx\phi_B(x,b=0)=\frac{f_B}{2\sqrt{2N_c}}\;,
\end{eqnarray}
and the variable $b$ conjugate to the parton transverse momentum
in the $B$ meson. The errors in the superscripts (subscripts) come
from $\omega_B=0.36$ ($\omega_B=0.44$) GeV. The range
$\omega_B=0.40\mp 0.04$ GeV, determined by a fit to the $B\to\pi$
form factor from light-cone sum rules \cite{TLS,CKL}, leads to the
$B\to K^*$ form factors $A_0=0.28^{+0.04}_{-0.03}$,
$A_1=0.26^{+0.04}_{-0.03}$, and $V=0.35^{+0.06}_{-0.04}$. The
above theoretical errors simply mean those arising from the
unknown $B$ meson wave function. It is clear that the polarization
fractions are insensitive to this source of uncertainties. There
are certainty other sources of theoretical uncertainties, whose
detailed investigation is not the focus of this work.



If our explanation is valid, the $B\to\phi K^*$ decays can be
classified into the same category as of $B\to \rho K^*$, for which
the subleading penguin annihilation and nonfactorizable
contributions render the polarization fractions deviate from the
naive counting rules based on kinematics. The only difference is
that the form factor ratio $A_0/A_1$ for the $B\to K^*$ transition
is smaller than that for the $B\to\rho$ transition. For example,
LFQCD gave $A_0/A_1=1.19$ for the former, and $A_0/A_1=1.27$ for
the latter \cite{CCH}. PQCD, using the set of $\rho$ meson
distribution amplitudes from \cite{BBKT}, gave $A_0/A_1=1.63$
\cite{TLS}. The PQCD results for the $B\to\pi\rho$ decays, to
which the $B\to\rho$ form factors are relevant, have been in good
agreement with the data \cite{LY00}. It implies that the values of
the $B\to\rho$ form factors yielded in PQCD are reasonable.
Therefore, the longitudinal polarization fraction $R_L$ of the
$B^+\to\rho^+ K^{*0}$ decay should be larger than that of $B\to
\phi K^*$. An explicit study of the $B\to\rho K^*$ modes in PQCD
is in progress. Furthermore, if our explanation is correct, the
modes governed by the $B\to K^*$ form factors, such as $B^+\to
K^{*+} K^{*0}$ and $B^0\to K^{*0} {\bar K}^{*0}$, must have low
$R_L\approx 0.6$. Note that the $B\to\phi K^*$ modes occur through
the $b\to s$ penguin, while the $B\to K^*K^*$ modes occur through
the $b\to d$ penguin. From the viewpoint of PQCD, the $B\to\omega
K^*$ decays, without exotic mechanism, show $R_L$ close to those
of $B\to\rho^0 K^*$. All the above predictions can be confronted
with the future polarization measurement of the $B\to K^*K^*$,
$\omega K^*$ decays.

At last, we compare the polarization fractions of the $B\to\rho
K^*$, $\omega K^*$ decays derived from the different approaches
with the $B\to \phi K^*$ data in Table~\ref{tab2}. The prediction
from the rescattering effect \cite{CDP,LLNS,CCS} is easily
understood: the $B\to \rho K^*$, $\phi K^*$ decays involve the
same $D_s^{(*)}D^{(*)}$ intermediate states, and their
polarization fractions are certainly almost equal. The relation
$R_\parallel(\phi K^*)\gg R_\perp(\phi K^*)$ is attributed to the
vanishing $R_\perp$ in the $B\to D_s^{*}D^{*}$ channels. The $b\to
sg$ transition contributes to the $B\to \omega K^*$, $\phi K^*$
modes, such that their polarization fractions are close to each
other \cite{HN}. Without the rescattering effect, the parallel and
perpendicular polarization fractions are expected to be similar.
This is also the case in PQCD \cite{CKL2} and in QCDF \cite{AK}.
The future data can provide an unambiguous discrimination among
these proposals. We did not list QCDF in Table~\ref{tab2}, because
both the relations $R_L(\rho^+ K^{*0}) > R_L (\phi K^{*0})$ and
$R_L(\rho^+ K^{*0}) \approx R_L (\phi K^{*0})$ are allowed within
the involved large theoretical uncertainty \cite{AK2}. Hence, it
is difficult to discriminate QCDF experimentally from the others.


\begin{table}[ht]
\begin{center}
\begin{tabular}{c c c}\hline \hline

PQCD&$R_L(\rho K^{*})>R_L (\phi K^{*})$ &
$R_\parallel(\phi K^{*}) \approx R_\perp (\phi K^{*})$\\
&$R_L(\omega K^{*}) \approx R_L(\rho^0 K^{*})$&\\
rescattering &$R_L(\rho K^{*}) \approx R_L
(\phi K^{*})$ &$R_\parallel(\phi K^{*}) \gg R_\perp (\phi K^{*})$\\
magnetic dipole &$R_L(\omega K^{*}) \approx R_L (\phi
K^{*})$&$R_\parallel(\phi K^{*}) \approx R_\perp (\phi K^{*})$\\
\hline \hline

\end{tabular}
\end{center}
\caption{Comparison of the $B\to\rho K^*$, $\omega K^*$
polarization fractions from the different proposals with the
$B\to\phi K^*$ data.}\label{tab2}
\end{table}

In this letter we have proposed a possible resolution to the
$B\to\phi K^*$ polarization puzzle within the Standard Model,
which was found to be nothing but the consequence of a smaller
$B\to K^*$ form factor $A_0$. Postulating $A_0\approx 0.3$, which
does not contradict to any existing measurement, we are able to
explain the data in the PQCD approach. This form factor value
leads to $R_L=0.84$ in FA (smaller than 0.92 corresponding to
$A_0=0.40$ in Table~\ref{tab5}). The penguin annihilation from the
$(S-P)(S+P)$ operators and the nonfactorizable contribution, which
can be estimated reliably in PQCD, then further bring $R_L$ down
to 0.59. The remaining concern is whether the asymptotic models in
Eqs.~(\ref{pk2})-(\ref{pk3s}) are allowed within theoretical
uncertainty. We have suggested to reanalyze these $K^*$ meson
distribution amplitudes appearing in the factorization formula for
$A_0$ in the framework of QCD sum rules.
Because of the unknown $A_0$, which is a source of QCD
uncertainty, it is still too early to claim any exotic mechanism
or new physics in the $B\to\phi K^*$ polarization data. We have
predicted that the penguin-dominated decays governed by the $B\to
K^*$ form factors, such as $B^+\to K^{*+} K^{*0}$ and $B^0\to
K^{*0} {\bar K}^{*0}$, should exhibit similar $R_L\approx 0.6$.
The comparison among the different proposals for resolving the
puzzle has been summarized in Table~\ref{tab2}, which can be
discriminated experimentally in the near future.


\vskip 0.5cm

I thank C.H. Chen for the numerical analysis, and H.Y. Cheng, C.K.
Chua, A. Kagan, and S. Mishima for useful discussions. This work
was supported by the National Science Council of R.O.C. under
Grant No. NSC-93-2112-M-001-014 and by the Taipei Branch of the
National Center for Theoretical Sciences of R.O.C..

\end{document}